\documentclass[superscriptaddress,showpacs,aps,twocolumn,prb,floatfix]{revtex4-1}
\usepackage{bm,amsmath,amssymb}
\usepackage{amssymb}
\usepackage{amsmath}
\usepackage{commath}
\usepackage{graphicx,bm}
\usepackage{verbatim}
\usepackage[latin1]{inputenc}
\usepackage[dvipsnames]{xcolor}
\usepackage[separate-uncertainty = true]{siunitx}
\usepackage[colorlinks=true,
            linkcolor = blue,
            urlcolor = RoyalBlue,
            citecolor = magenta]
            {hyperref}

\begin{document}

\title{Thermoelectricity in Quantum-Hall Corbino Structures}

\author{Mariano Real}
\affiliation{Instituto Nacional de Tecnolog\'{\i}a Industrial, INTI and INCALIN-UNSAM, Av. Gral. Paz 5445, (1650) Buenos Aires, Argentina}
\author{Daniel Gresta}
\affiliation{International Center for Advanced Studies, ECyT-UNSAM,  25 de Mayo y Francia, 1650 Buenos Aires, Argentina}
\author{Christian Reichl}
\affiliation{Solid State Physics Laboratory, ETH Z\"urich, CH-8093 Z\"urich, Switzerland}
\author{J\"urgen Weis}
\affiliation{Max-Plack-Institut f\"ur Festk\"orperforschung, Heisenbergstrasse 1, D-70569 Stuttgart, Germany}
\author{Alejandra Tonina}
\affiliation{Instituto Nacional de Tecnolog\'{\i}a Industrial, INTI and INCALIN-UNSAM, Av. Gral. Paz 5445, (1650) Buenos Aires, Argentina}
\author{Paula Giudici}
\affiliation{INN CNEA-CONICET, Av. Gral. Paz 1499 (1650) Buenos Aires, Argentina}
\author{Liliana Arrachea}
\affiliation{International Center for Advanced Studies, ECyT-UNSAM, 25 de Mayo y Francia, 1650 Buenos Aires, Argentina}
\author{Werner Wegscheider}
\affiliation{Solid State Physics Laboratory, ETH Z\"urich, CH-8093 Z\"urich, Switzerland}
\author{Werner Dietsche}
\affiliation{Solid State Physics Laboratory, ETH Z\"urich, CH-8093 Z\"urich, Switzerland}
\affiliation{Max-Plack-Institut f\"ur Festk\"orperforschung, Heisenbergstrasse 1, D-70569 Stuttgart, Germany}

\begin{abstract}
We measure the thermoelectric response of Corbino structures in the quantum Hall effect regime and compare it with a theoretical analysis. The measured thermoelectric voltages are qualitatively and quantitatively simulated based upon the independent measurement of the conductivity indicating that they originate predominantly from the electron diffusion.  Electron-phonon interaction does not lead to a phonon-drag contribution in contrast to earlier Hall-bar experiments. This implies a description of the Onsager coefficients on the basis of a single transmission function, from which both thermovoltage and conductivity can be predicted with a single fitting parameter. It furthermore let us predict a figure of merit for the efficiency of thermoelectric cooling which becomes very large for partially filled Landau levels (LL) and high magnetic fields.
\end{abstract}

\maketitle

\section{Introduction}

The quantum Hall effect (QHE) which occurs in two-dimensional electron systems (2DES) exposed to quantizing magnetic fields is one of the most prominent examples of the synergy between fundamental physics and quantum technologies \cite{vK}. It is topological in nature and intrinsically  related to exotic properties of matter, like fractionalization and non-abelian statistics \cite{fractio,laughlin,jain}. At the same time,  these complex properties are precisely the reason for its robustness and appeal for practical applications. It is nowadays at the heart of the definition of the electrical metrological standards \cite{gerster}, while it is also a promising platform for the development of topological quantum computation \cite{topocomp}. 

\begin{figure}
    \centering
    \includegraphics[width=\columnwidth]{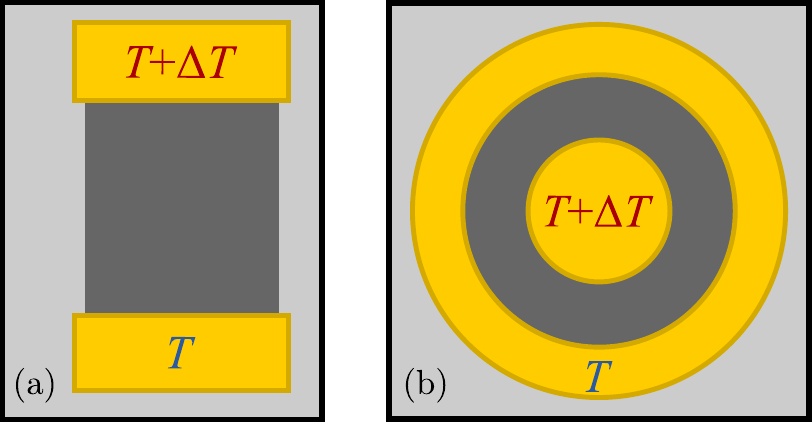}
    \caption{The two sample designs to investigate thermoelectric effects, Hall-bar (a) and Corbino (b). The dark gray areas are the 2DES. The hot and the cold contacts for measuring the thermovoltage are at two ends of the rectangular shaped Hall-bar. For the Corbino, the hot contact is in the center of the 2DES which is surrounded by the cold one.}
    \label{fig:BasicDesigns}
\end{figure}

Measurements of the entropy would be of great importance in verifying the theoretically expected quantum states, particularly of the non-abelian ones. One possibility to access entropy in a 2DES is measuring thermoelectricity \cite{halpe,barlas}. But, although thermoelectricity had been studied both experimentally and theoretically since the discovery of the integer QHE, \cite{Fletcher88,Fromhold93,therpow8} it had  not been possible to reconcile the experimental results obtained with Hall-bars,  Fig.~\ref{fig:BasicDesigns} (a), with theories based upon electron diffusion. The overwhelming effect of phonon-drag was invoked as one reason \cite{Fletcher88,Fromhold93,Maximov_2004}, but more recently inherent  problems connected with the topology of the Hall-bar geometry, affecting both phonon-drag and electron diffusion, have been realized \cite{barlas}. The longitudinal thermopower (or Seebeck coefficient) measured along a Hall bar resembles closely the longitudinal resistance, both in the phonon-drag and in the diffusive regime while an oscillating behavior with sign changes was expected by the theory. It was suggested that the longitudinal thermopower could be measured correctly in Corbino geometry, Fig.~\ref{fig:BasicDesigns} (b), where due to the circular geometry, the thermal bias is applied radially, hence, any thermal and electrical transport is induced along the radial direction \cite{dolgo} and the transport takes place through the bulk.

Early thermopower experiments in Corbino geometry failed to observe the expected sign-changing behavior\cite{therpow10}. It was reported however in other experimental works using the Corbino geometry \cite{kobaya}. The latter used rf-heating of the 2DES to produce the temperature gradient directly in the 2DES claiming that no phonons are involved in the measured thermopower.

In this article we report Corbino thermopower measurements in the QHE regime by using a conventional heater in the center of the device. This way, a temperature gradient is set up in both the substrate and the 2DES. Results at temperatures from about 300 mK to 2 K are presented and compared with theoretical results where both electrical conductance and thermopower are modeled based upon the same transmission function. A very good agreement over our range of magnetic fields and temperatures is found. This demonstrates that the substantial disagreement which was typical for Hall bars can be removed by using the Corbino topology. In contrast to Hall-bar studies, it is not necessary to consider phonon-drag in the theory.

\section{Experimental details}
\subsection{Setup}

\begin{figure}
    \centering
    \includegraphics[width=\columnwidth]{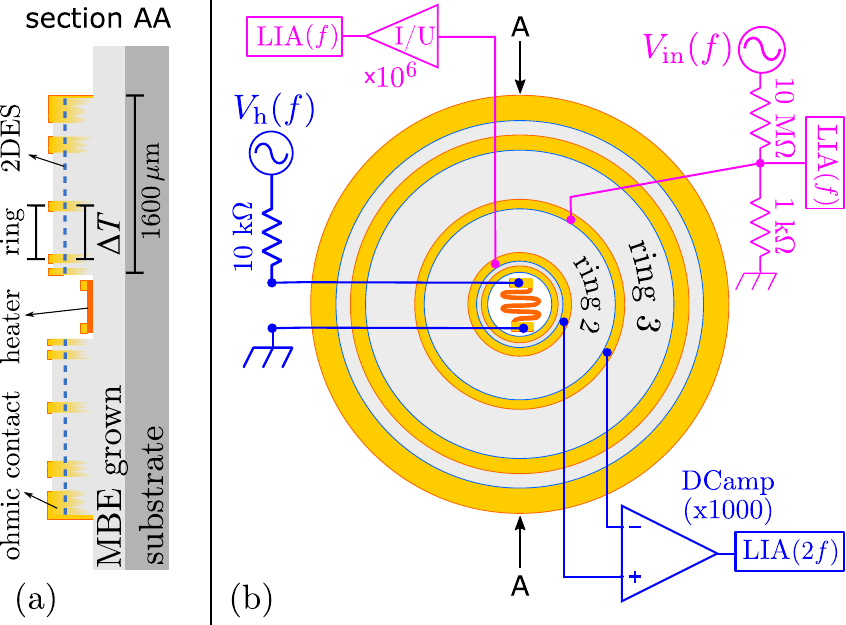}
    \caption{Scheme of experimental setup. (a) Cross-section of the sample, notice that the heater element is over the substrate outside the 2DES. (b) Measurement configurations for the conductance and the thermovoltage are shown in light-grey (magenta online) and black (blue online), respectively. LIA denotes lock-in amplifier. The two type of experiments were done in separate runs. Only two of the four Corbino rings are labeled in the figure.}
    \label{fig:fig1}
\end{figure}

Our setup is sketched in Fig.~\ref{fig:fig1}. An AuPd thin-film heater is inserted in the center of the Corbino samples and heated with an
AC current with a frequency $f$ of a few Hertz producing a temperature oscillation of 2$f$. In this way, a radial thermal gradient is induced between the center and the external edge of the sample, which is assumed to be close to the temperature $T$ of the bath. The device used here consists of five concentic ohmic contact rings  with diameters ranging from \SIrange{.4}{3.2}{\milli\meter} made by alloying Au-Ge-Ni into the 2DES structure forming four independent Corbino rings. Under the heater and outside of the rings the 2DES is removed. It is assumed that the local temperature over the 2DES follows the one of the underlying GaAs substrate. This was already verified by Chickering et al. \cite{chick1,chick2} down to much lower temperatures than the ones used here. We neglect possible anisotropies in the heat conductivity of the substrate due to the ballistic nature of the phonons because these become only relevant if the dimension of the heater and the contacts are much smaller than the substrate thickness \cite{Wolfe}.

The four Corbino rings of this device allow not only to measure thermopower at four different radial distances from the heater but also to determine, in a different experiment, the temperatures at the different ring positions. This can be done by using the conductances of the four Corbinos as thermometers. Measuring the conductance as function of bath temperature without any heat applied is used for calibration. With the heater on, the temperatures at the different rings can be measured.

As a response to the thermal bias between the center and the edge of the sample, charges diffuse across the Corbino ring which is compensated by generating a voltage with frequecy 2$f$ between the inner and the outer circumferences. The sign and the magnitude of this thermovoltage is determined by the transmission function as discussed in the theory section.
The thermoelectric response in this device is much simpler than  the one in the  Hall bar geometry, where the transport takes place along longitudinal and transverse directions with respect to the applied biases. In fact, in the Corbino geometry, the thermovoltage develops along the direction of the temperature bias. 

The samples were grown by molecular beam-epitaxy on GaAs wafers having a
single 2DES located in a 30 nm wide quantum well with Si-doped doped layers on both sides. Data from two samples from two wafers, A and B are presented here. Separate test pieces from these wafers in van-der-Pauw geometry had mobilities of \SI{21e6}{\square\centi\meter\per\volt\per\second} and \SI{18e6}{\square\centi\meter\per\volt\per\second} at electron densities of $n_e = \SI{3.06e11}{\per\square\centi\meter}$ and  $n_e = \SI{2.0e11}{\per\square\centi\meter}$, respectively, measured at \SI{1.3}{\kelvin} in the dark. 
 
The Corbino samples were glued in a standard commercial ceramic
holder with gold-plated pins and base and a \SI{3}{\milli\meter} diameter hole drilled in
the middle to reduce thermal contact to the samples.
The measurements were performed in vacuum in a $^3$He cryostat equipped with a \SI{14}{\tesla} magnet being able to achieve a base temperature of \SI{250}{\milli\kelvin}. 

Fig.~\ref{fig:fig1} also shows the configurations used for the measurements of the conductance (light-gray, magenta online) and of the thermovoltage (black, blue online).  The conductance $G$ was measured by applying an AC voltage through a voltage divider and measuring the current with an amplifier (IUAmp). 
Thermovoltage $V_{tp}$ was measured in separate runs by passing an AC current of frequency $f$ to the central heater having a resistance of  about \SI{650}{\ohm}. The thermopower induced in the sample was measured by using a $\times 1000$ differential DC voltage amplifier (DCamp). The input impedance of this amplifier must be very high because the internal resistance of the Corbino device diverges in the quantum-Hall states. We used an amplifier with an iput impedance of about \SI{1}{\tera\ohm} \cite{marki2015}. Very little frequency dependence of the thermovoltage was found between $f$ = \SI{3}{\hertz} and \SI{100}{\hertz}. Most measurements were done at  $f$ = \SI{13.8}{\hertz}.
To avoid effects of time-dependent magnetic fluxes, the waiting time for each data point was set to a few seconds to guarantee the stabilization of the magnetic fields at a constant value.

\subsection{Thermovoltage measurement}
\begin{figure}
    \centering
    \includegraphics[width=\columnwidth]{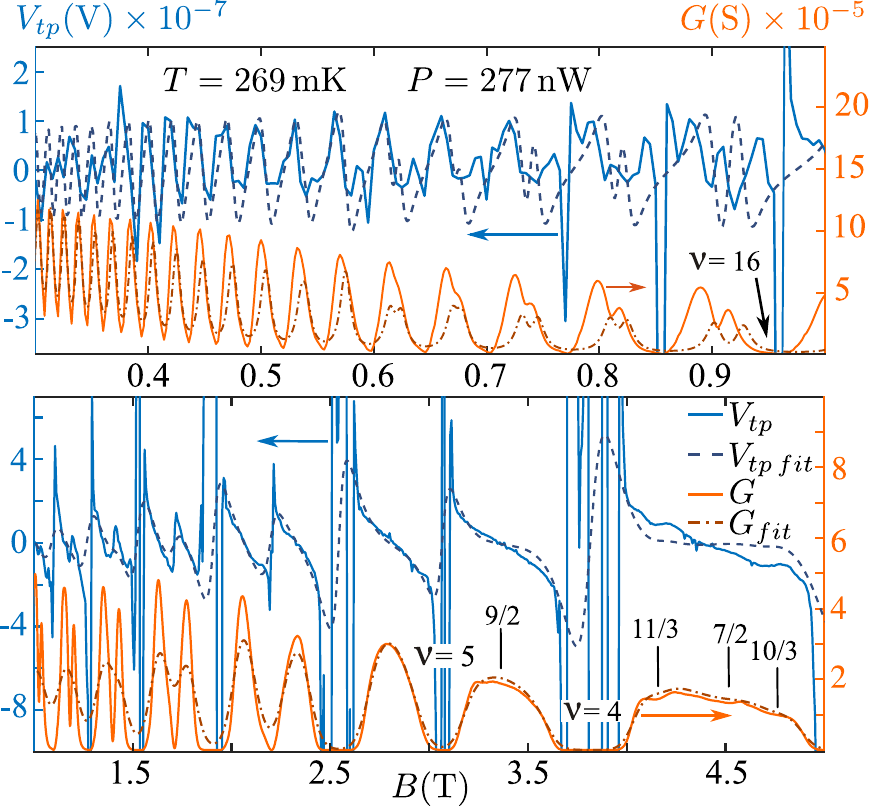}
    \caption{Conductance $G$ and thermovoltage $V_{tp}$ as a function of the magnetic field $B$ for the ring 2 in Fig.~\ref{fig:fig1} at temperature $T$ with power $P$ supplied at the heater.
        Experimental data is plotted in solid lines. Theoretical (dashed) plots  are based on the calculation of Eq.~(\ref{onsa}) with the respective inferred transmission functions.}
    \label{fig:fig2}
\end{figure}

Experimental results for both the thermal voltage (solid blue) and the conductance (solid orange) of sample A are shown in Fig.~\ref{fig:fig2} at a base temperature $T$ of  269 mK and an average heater power $P$ of 277 nW. The magnetic field is swept from 0.3 Tesla to 5 Tesla. The conductance shows the typical Shubnikov-de-Haas (SdH) oscillations with the spin splitting becoming visible at about 0.9 Tesla and the conductance minima aproaching zero at even filling factors less than 20. The thermovoltage  $V_{tp}$ shows numerous features. At small magnetic fields, it oscillates with a similar periodicity as the conductance changing sign both at the conductance maxima and minima. At higher magnetic fields additional features appear in the regions of the conductance minima which become very signifcant and chaotic at even larger magnetic fields where the conductance minima are wider. Between the conductance minima, the thermovoltage $V_{tp}$ now changes to saw-tooth like behavior, still changing signs at both the maxima and the minima of the conductance. Such sign changes had not been observed in the earlier Hall-bar experiments but were already seen in the previous rf-based Corbino experiments\cite{kobaya} and had been expected theoretically \cite{barlas}.

\begin{figure}
    \centering
    \includegraphics[width=\columnwidth]{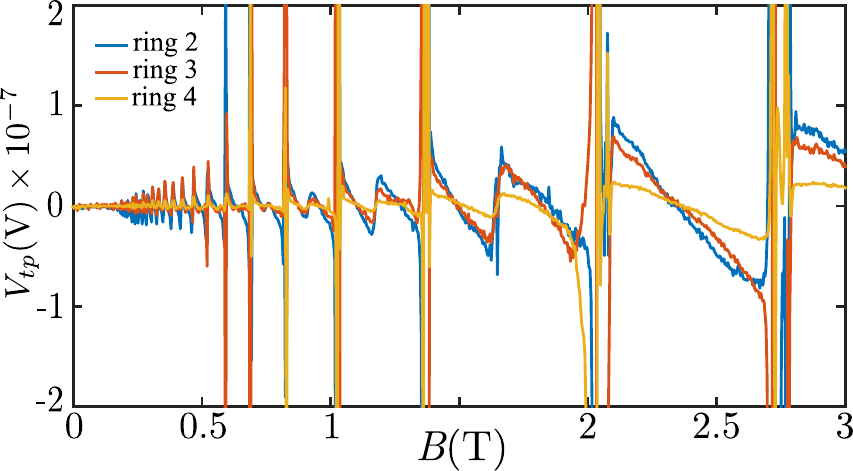}
    \caption{$V_{tp}$ response of sample B at different rings at
        a bath temperature of 600 mK and a heater power of 213 nW.   ring 2 and ring 3 present a greater temperature gradient and hence a larger voltage response than  ring 4.}
    \label{fig:fig2SM}
\end{figure}

We have observed the sign change behavior in a similar way on several samples with different densities and mobilities. Data of the sample B are presented in Fig.~\ref{fig:fig2SM} showing $V_{tp}$ measured across the three different outer Corbino rings at a temperature of 600 mK. The oscillatory behavior of $V_{tp}$ is again clearly visible as are the large signals in the regions which correspond to the conductance minima.

The large signals have not been reported before. These are no spurious signals. They are reproducible, they persist if the magnetic field is stopped and kept constant or if the sweep direction is reversed. They are definitely thermally induced signals and are not produced by an electromagnetic crosstalk. The large signals vanish by applying a dc current ontop of a square AC current. This leads to a constant heating and thus a vanishing temperature oscillation but would leave any suspected crosstalk unchanged. We will speculate at the end of this paper about possible origins of the large signals. 
 
In the following we will concentrate on the analysis of the thermal voltages outside of the SdH minima. We show that the magnetic field trace of both $V_{tp}$ and conductance $G$ can be fitted using only charge diffusion. The same transmission function based upon model Landau levels is used for calculating both conductance and thermovoltage. The only fitting parameter will be the temperature gradient. The resulting fits are already shown in  Fig.~\ref{fig:fig2} as dashed lines.

\section{Thermoelectric response} 
\subsection{Onsager coefficients}
We consider the Corbino geometry in Fig.~\ref{fig:BasicDesigns} (b) to describe the thermoelectric transport. The Corbino ring acts as a conductor in radial direction between hot and cold reservoirs with a temperature bias of $\Delta T$.
 In linear response, the corresponding charge and heat currents for small  $\Delta T$ and bias voltage $V$
 can be expressed as \cite{casati}
\begin{equation}\label{thermo}
\left(
\begin{array}{c}
I^C/e \\
I^Q 
\end{array}
\right)  =  \left(
\begin{array}{cc}
{\cal L}_{11} &{\cal  L}_{12}  \\
{\cal L}_{21} &{\cal L}_{22} 
\end{array}
\right) 
\left(
\begin{array}{c}
X_1\\
X_2
\end{array}
\right),
\end{equation}
where  $X_1=eV /k_B T$ and $X_2= \Delta T/k_B T^2$ and $\hat{\cal L}$ is  the Onsager matrix. 
The electrical and thermal conductances are, respectively, $G=e^2 {\cal L}_{11}/T$, and 
$\kappa=\mbox{Det}{\hat{\cal L}}/\left(T^2 {\cal L}_{11} \right)$. $S={\cal L}_{12}/{\cal L}_{11}$ defines  the Seebeck and 
$\Pi={\cal L}_{21}/{\cal L}_{11}$ is the Peltier coefficient. 
For ballistic or diffusive transport  ${\cal L}_{ij}$ depends only on the quantum dynamics of the electrons  in the presence of the magnetic field and the disorder of the sample. They are  described by 
  a transmission function ${\cal T}(\varepsilon)$,
\begin{equation}\label{onsa}
{\cal L}_{ ij}= - T \int \frac{d \varepsilon}{h} \frac{\partial f (\varepsilon)}{\partial \varepsilon} \left(\varepsilon-\mu \right)^{i+j-2} {\cal T}(\varepsilon),
\end{equation}
where  $f(\varepsilon)=1/(e^{(\varepsilon-\mu)/k_B T}+1)$ is the Fermi distribution function, $\mu$ is the chemical potential and $T$ is the temperature of the electrons. In the presence of disorder and absence of electron-electron interactions ${\cal T}(\varepsilon)$ was originally calculated by Jonson and Girvin \cite{jonson}. At high temperatures, electron-phonon interaction gives rise to an additional component to the transport coefficients ${\cal L}_{ij}$.

\subsection{Conductance and thermovoltage.} Our goal is to accurately describe the electronic component of the Onsager coefficients obtained from the experimental data. 
At first we measure the conductance $G(B)$ as a function of the applied magnetic field $B$.
The thermovoltage $V_{tp}$ is measured separately and corresponds to the voltage for which $I^C=0$ in Eq. (\ref{thermo}),
\begin{equation}\label{vtp}
V_{tp}(B)= - S(B) \frac{\Delta T}{T}.
\end{equation}
Here $S(B)$ is the Seebeck coefficient as a function of the magnetic field, $T$ is the temperature of the bath (cold finger in our case) and $\Delta T$ is the temperature difference between the two contacts of the Corbino ring under investigation.
From the data of $G(B)$ we infer the transmission function ${\cal T}(\varepsilon)$ entering Eq. (\ref{onsa}). Given ${\cal T}(\varepsilon)$,
we can evaluate the electrical component of the other Onsager coefficients, in particular ${\cal L}_{12}(B)$. Through Eq. (\ref{vtp}),
this leads to a theoretical prediction for the behavior of $V_{tp}(B)$ resulting from the electrical transport,  which can be directly contrasted with the experimental data.  

There are two regimes to be considered for the calculation of ${\cal T}(\varepsilon)$: (i) At low magnetic fields, where the different Landau levels are not clearly resolved, we calculate the transmission function with the model introduced in Ref. \cite{jonson,barlas}. The latter is based on a single-particle picture for the 2DES in the presence of a magnetic field and elastic scattering introduced by impurities. (ii) For higher magnetic fields, where the different filled LL are clearly distinguished, and separated by a gap, we use the fact that in the limit of $T \rightarrow 0$,  Eq.~(\ref{onsa}) leads to ${\cal T}(\mu) \sim G(\mu )/e$.

\subsection{Transmission function}
\subsubsection{Low magnetic field}
Here we consider the transmission function\cite{barlas,jonson} 
\begin{eqnarray}\label{girvin}
{\cal T}(\varepsilon)&=& \Lambda \sum_{n,\sigma}  \frac{\left(n+1 \right) \omega_c^2 }{8 \pi h}  A_{n,\sigma}(\varepsilon) A_{n+1,\sigma}(\varepsilon),
\end{eqnarray}
where $\Lambda$ is a geometric factor relating the conductance to the conductivity, while $A_{n,\sigma}(\varepsilon)=\mbox{Im}\left[G_{n,\sigma}(\varepsilon)\right]$, being 
$G_{n,\sigma}(\varepsilon)= \left[ \varepsilon - \varepsilon_{n,\sigma} - \Sigma(\varepsilon) \right]^{-1}$  the Green function calculated within the self-consistent Born approximation.
$\varepsilon_{n,\sigma}=\hbar (n+1/2) \omega_c \pm \mu_B B/2 $ is the energy of the Landau levels, including the Zeeman splitting, with $\pm$ corresponding, respectively, 
to $\sigma=\uparrow, \downarrow$. 
Here,
$\mu_B$ is the Born magneton,  $\omega_c= e B/m^* $ is the cyclotron frequency, and  $m^*=0.067m_e$ is the effective mass of the electrons in the structure and $m_e$ is the electron mass.
The effect of disorder due to impurities introduces a widening $\Gamma$ in the Landau levels, which is accounted for the self-energy 
$\Sigma(\varepsilon)=(\omega-\varepsilon_L)/2 - i \Gamma \sqrt{1- (\varepsilon-\varepsilon_L)^2/(4\Gamma^2)}$. Here $\varepsilon_L$ is the energy of the Landau level which is closest to $\varepsilon$.
This model has two fitting parameters: $\Lambda$ and $\Gamma$, which we adjust to fit the data of the conductance $G$, through Eq.~(\ref{onsa}).
This model fails to reproduce $G(B)$ for high magnetic fields ($B > \SI{1}{\tesla}$). Thus a different model has to be used in this regime.

\subsubsection{High magnetic field}
For higher magnetic fields, satisfying $k_B T \ll \hbar \omega_c$, and $\Gamma \ll \hbar \omega_c  $,  we can  infer the transmission function more efficiently
from the behavior of the conductance within a range of magnetic fields in the neighborhood of a given filling fraction $\nu$. Notice that in the limit of $T \rightarrow 0$, the derivative of the Fermi function entering Eq.~(2) of the main text, has the following behavior,
$-\partial f(\varepsilon)/\partial \varepsilon \rightarrow \delta(\varepsilon-\mu)$. Therefore, for low temperatures, such that  $k_B T \ll \hbar \omega_c$, we have
\begin{equation}\label{high}
{\cal T}(\mu_{\nu} ) \sim \frac{G(\mu_{\nu} )}{e}, \;\;\;\; \mu_{\nu}= \frac{\hbar e B}{2 m^* }, \;\;\; B_{\nu+1} < B < B_{\nu},
\end{equation}
where $B_{\nu}=n_e h/(e\nu)$  is the magnetic field corresponding to the filing fraction $\nu$, while $\mu_{\nu} $ is the Fermi energy for the range of $B$ within two consecutive integer filling factors.

\section{Results}
\subsection{Thermoelectric response}

Results for the conductance and the thermovoltage are shown in Fig.~\ref{fig:fig2} for the temperature \SI{269}{\milli\kelvin}. The experimental data for $G$ and $V_{tp}$ within the regime of low magnetic field is shown in the upper panel of the figure along with the theoretical description based on the transmission function of Eq.~(\ref{girvin}). 

In the case of high magnetic field, shown in the lower panel, the theoretical description was based in the transmission function  of Eq.~(\ref{high}).
Given ${\cal T}(\varepsilon)$, we calculate  the Onsager coefficients of Eq.~(\ref{onsa}) and the Seebeck  coefficient $S={\cal L}_{12}/{\cal L}_{11}$.  
The ratio $\Delta T/T$ has been adjusted in order to fit the experimental measurements with Eq.~(\ref{vtp}).    The estimates for the temperature bias were $\Delta T = \SI{1}{\milli\kelvin}$ and \SI{1.08}{\milli\kelvin}, for low and high magnetic fields, respectively.  Overall,  in particular for high magnetic fields, the agreement between experiment and theory is excellent within the range of $B$ corresponding to partially filled LL, for which $G \neq 0$. 
\begin{figure}
    \centering
    \includegraphics[width=\columnwidth]{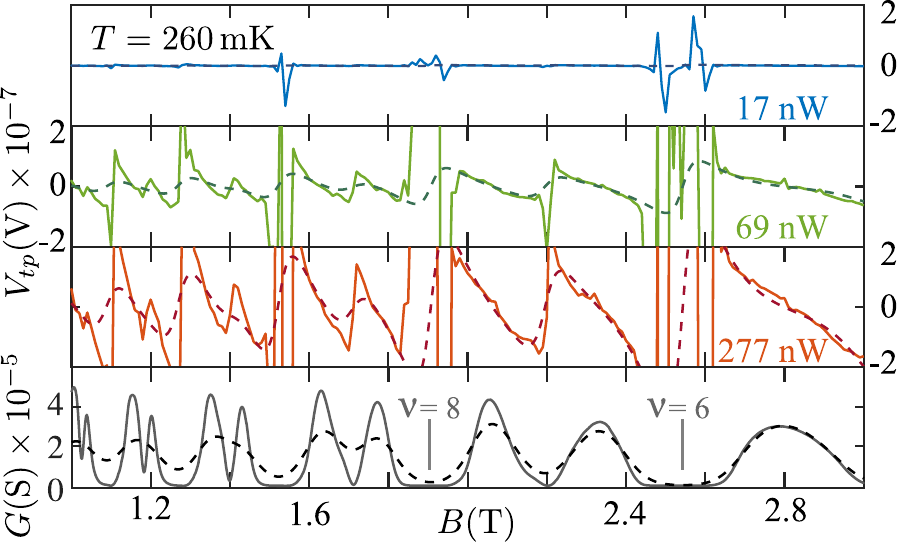}
    \caption{Thermovoltage $V_{tp}$  for a fixed temperature and different powers $P^{\prime}$
        applied at the heater, assuming $\Delta T (P^{\prime}) = P^{\prime}/P \; \SI{1.08}{\milli\kelvin}$.
        $P$ and other details are the same as in  Fig.~\ref{fig:fig2}.}
    \label{fig:fig3}
\end{figure}

\begin{figure}
	\centering
	\includegraphics[width=\columnwidth]{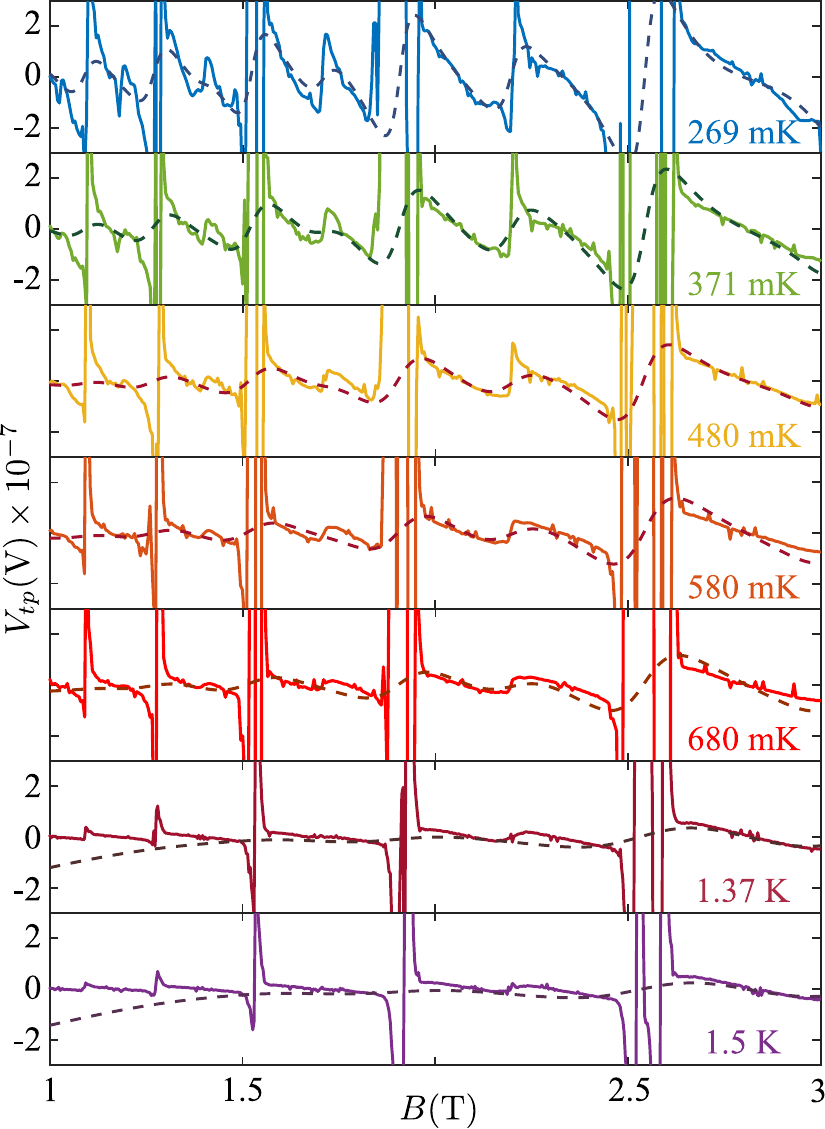}
	\caption{Thermovoltage $V_{tp}$, as function of the magnetic field for different temperatures.   In the case of \SIrange{269}{680}{\milli\kelvin} a power of \SI{277}{\nano\watt} was used, while for \SIrange{1.37}{1.5}{\kelvin} the heater power was \SI{433}{\nano\watt}. Other details are the same as in previous Figs. The scale for $V_{tp}$  is the same in all panels.}
	\label{fig:fig4}
\end{figure}
Taking into account the good agreement between the experimental and theoretical estimates of the temperature difference $\Delta T$ found in the analysis of the data of Fig.~\ref{fig:fig2},
we now analyze the relation between the electrical power supplied at the heater and $\Delta T$.  In Fig.~\ref{fig:fig3} we show experimental data for the thermovoltage at a fixed temperature and different heater powers. We have assumed a linear dependence between these quantities. Therefore, we have fitted the experimental data with the same Seebeck coefficient
$S(B)$ calculated for Fig.~\ref{fig:fig2} and the following values of the temperature difference, 
$\Delta T (P^{\prime}) = P^{\prime}/P \; \SI{1.08}{\milli\kelvin}$,  being $P^{\prime}$ the power corresponding to the experimental data and $P$ the power used in the data of Fig.~\ref{fig:fig2}. We see a very good agreement between the theoretical prediction and the experimental data.
In Fig. \ref{fig:fig4} we discuss the evolution of $V_{tp}$ as the temperature grows, focusing on the high magnetic field region. The experimental data is presented along with the theoretical prediction obtained by following the same procedure of the previous Figs, and taking into account
the linear dependence of $\Delta T$ with $P$ explained in Fig.
\ref{fig:fig3}. 
The agreement between the theoretical predictions and the experimental data for magnetic fields corresponding 
to partially filled LL within a wide range of temperature is overall very good, improving as the  temperature decreases.

\subsection{Theoretical estimate of $\Delta T$}
\begin{figure}
	\centering
	\includegraphics[width=\columnwidth]{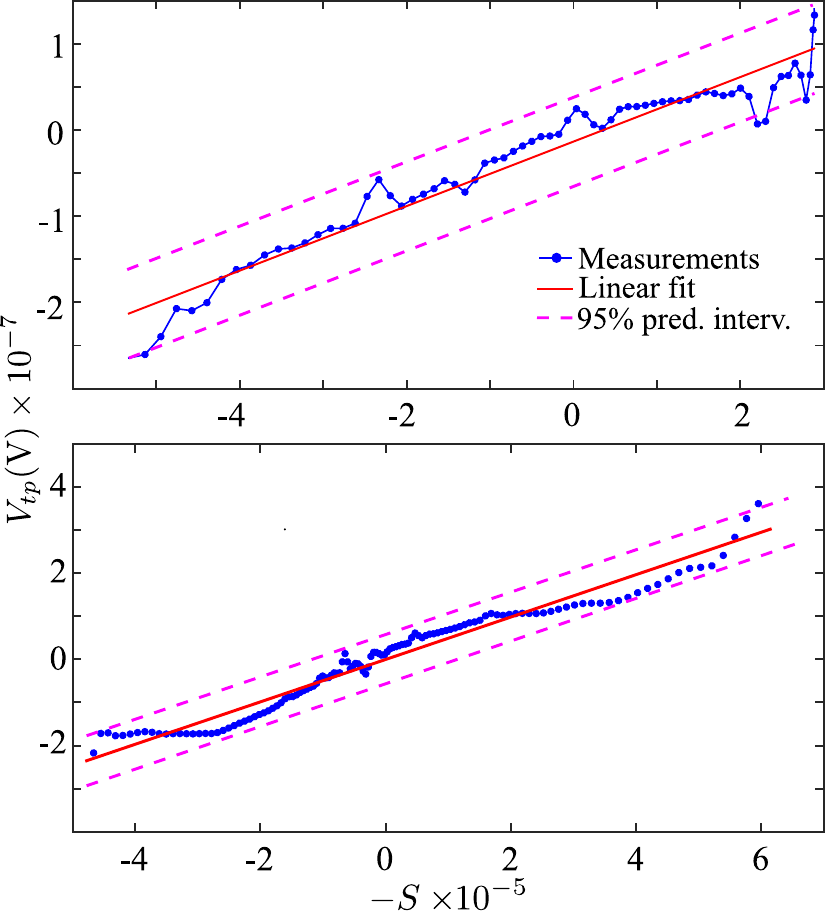}
	\caption{ Measured $V_{tp}$ signal {\em vs} calculated $-S$ 
		within the range of fields $B =$~\SIrange{2.21}{2.46}{\tesla} (upper panel) and $B =$~\SIrange{2.625}{3}{\tesla} (lower panel).
		The slope of this relation is $\Delta T/T$. }
	\label{fig:fig6SM}
\end{figure}

From the behavior of the conductance we can infer  the transmission function ${\cal T}(\varepsilon)$ as explained before, from where we can calculate the Onsager coefficients ${\cal L}_{ij}$. We recall that 
the thermovoltage is defined in Eq.~(\ref{vtp}). Given the calculation
of  $S(B)$,   we need to adjust the parameter $\Delta T/T$ in order to fit the data. Since the latter enters as the slope in the linear function $V(S)$, we analyze plots of the measured
$V_{tp}$ {\em vs}  the calculated $S$ for values of $B$ within which the Landau levels are partially filled and we fit a linear function to obtain the slope.
Examples are shown in Fig. \ref{fig:fig6SM} for two different Landau levels, with bath temperature $T= \SI{269}{\milli\kelvin}$ and a power of \SI{277}{\nano\watt}.

The corresponding fits cast $\Delta T= \SI{1.01(6)}{\milli\kelvin}$ in the region from $B =$~\SIrange{2.21}{2.46}{\tesla} (upper panel), and 
$\Delta T = \SI{1.33(6)}{\milli\kelvin}$ in the region from $B =$~\SIrange{2.625}{3}{\tesla} (lower panel), with uncertainties 
corresponding to a 95\% confidence probability.
Also notice that the intercept, which was taken as a free parameter of the regression is zero within the error in both cases.

The data of the other sample measured at 600 mK shown in Fig.~\ref{fig:fig2SM} can be analyzed similar for the several rings.
 For the ring 2 we obtain $\Delta T = \SI{60(3)}{\micro\kelvin}$,
  while for the ring 3 we get $(\Delta T = \SI{110(10)}{\micro\kelvin}$  
and for ring 4 $\Delta T = \SI{74(50)}{\micro\kelvin}$.  
 These values are considerably lower than the ones of the sample A measured at  \SI{269}{\milli\kelvin} . The reason is that 
the thermal conductivity of the substrate increases with $T^3$ and at \SI{600}{\milli\kelvin} the temperature gradient will be nearly 10 times smaller and, correspondingly, a much smaller temperature difference is expected at the higher temperature.

\subsection{Experimental estimate of $\Delta T$}

The exact determination of  $\Delta T$ in a Corbino device turns out to be challenging. The reason is the high phonon-heat conductivity in the GaAs substrate leads to small temperature gradients between the center and the edges of the sample. One consequence is that the thermal resistance from the sample to the ceramic carrier can no longer be neglected. Using the temperature dependent conductance of the Corbino rings as thermometers, we were nevertheless able to make estimates. The innermost and the outermost rings in Fig.~\ref{fig:fig1} were used for this measurement. The conductance minimum at filling factor 9 in Fig.~\ref{fig:fig2} was used because it showed a pronounced temperature dependence. Actual temperatures at these rings were found by comparing the respective  conductances with the heater on and off. The temperature rise at both rings could be close to \SI{50}{\milli\kelvin} with the cryostat at \SI{269}{\milli\kelvin} and heater powers reaching 300 nW. The temperature difference between the two rings was found to be less than \SI{9}{\milli\kelvin}  at the highest power. This number is only an estimate because the precision and the reproducibility of the calibration procedure was limited by the temperature control of the cryostat.

The determination of $\Delta T$ could have been improved by thinning the sample and thereby decreasing its thermal conductance. We have done this for several samples but the thermovoltage data as function of magnetic field became erratic. We suspect that the thinning led to inhomogeneities in the 2DES making the $V_{tp}$ measurements useless. Doping of the substrate with Cr would be another way to decrease the thermal conductivity.

Alternatively, we estimated the temperature profile using literature values of the thermal conductivity $\kappa$ . From \cite{chick1} we deduced a $\kappa$ of about 0.01 W/mK at \SI{300}{\milli\kelvin}. Using the simulation software Comsol a temperature difference of \SI{2.5}{\milli\kelvin} between the center and edge of our sample  was found from the heat-flow equation. This would lead to a temperature difference across ring 2 of about \SI{1}{\milli\kelvin} in good agreement with the used fit values.

\begin{figure}
	\centering
	\includegraphics[width=\columnwidth]{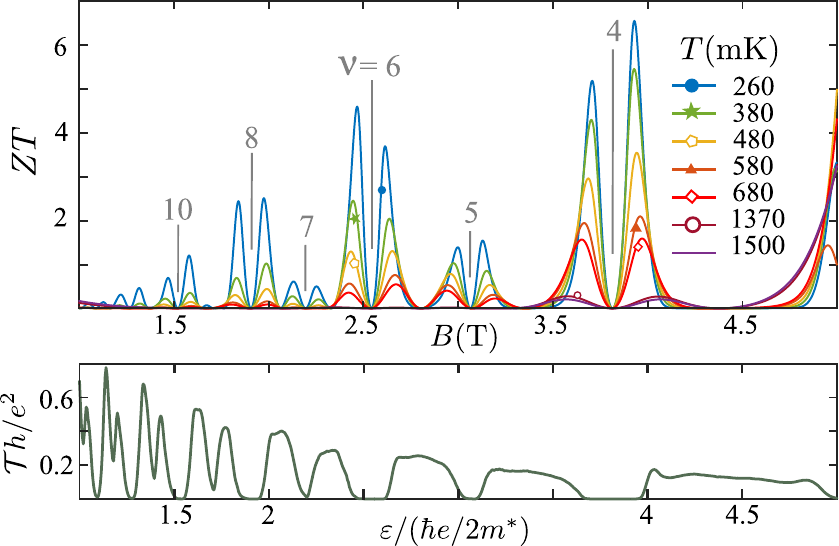}
	\caption{Bottom: Transmission function ${\cal T}(\varepsilon)$. Top: Electron contribution to the figure of merit $ZT$.}
	\label{fig:fig5}
\end{figure}

\section{Thermoelectric performance}

The quality of the thermoelectric performance, i.e.
the efficiency (for a heat engine), 
 or coefficient of performance (for a refrigerator) has found great interest in recent years, particularly in the context of the ballistic transport along edge channels and nano sized devices\cite{rafa,janine,daniel,hof,peter,vanuci,enhan,fabio2,fu}.
 We find that the theoretical performance of a Corbino device in the Quantum-Hall regime is surprisingly large being comparable to the highest predicted values in devices based upon ballistic transport. The performance is parameterized as the figure of merit \cite{casati}, $ZT = {\cal L}_{21}^2/\mbox{Det}{\hat{\cal L}}$, or $S^2$ times the ratio of the electric and thermal conductivity.
 
The optimal Carnot efficiency/coefficient of performance is achieved for $ZT \rightarrow \infty$.
 The highest reported values in real, usually semiconducting materials are between 
 $1 \leq ZT\leq 2.7$ \cite{casati,he} while optimistic theoretical predictions in the ballistic edge-channel regime are
 $ZT \sim 4$ \cite{sebas1} or lower.
 In Fig.~\ref{fig:fig5} we show the transmission function ${\cal T}(\varepsilon)$ used for the Corbino in this work to fit the experimental data of Fig.~\ref{fig:fig2} within the high-magnetic field regime. We see that the sequence of sharp features at the LL realize energy filters, leading to large values of  $ZT \sim 6$. Thus, diffusive transport across the bulk of a Corbino device has a potentially higher performance than the evisioned edge-channel devices.  Fig.~\ref{fig:fig5} suggests that even higher $ZT$ values should be possible at lower temperatures.
We stress that this analysis is based on the assumption that the main contribution to the thermoelectric and thermal transport is due to the electrons.
Phononic thermal transport in the substrate would tend to decrease the performance but would die out at even lower temperatures with $T^3$ while the figure of merit would probably increase. Thus one could envision that the Corbino device could be used as a thermoelectric cooler in the low mK regime for specific purposes. Replacing the heater by the object to be cooled could already be sufficient to form a realistic device.

\section{Conclusions}
We analyzed the thermoelectric response of a Corbino structure in the quantum Hall regime. For partially filled Landau levels, we found an excellent agreement between the experimental data and the
theoretical description based on the assumption that the thermoelectric response originates in the diffusion of electrons while electron-phonon drag does not influence the thermovoltage in temperature range from 300 mK to 2 K. Clearly, the electron-phonon interaction does not vanish in the Corbino geometry, but the transfer of momentum from the phonons to the electrons does not lead to a measurable voltage. Actually, it had been already noted long time ago that the 
contribution of the phonon-drag mechanism to the thermoelectric coefficient ${\cal L}_{12}$ should be zero in the heat-flow direction \cite{Fromhold93} which is, simply put, a consequence of the Lorentz force. It appears that only in Corbino devices the vanishing contribution of phonon-drag is reflected in the thermovoltage measurement. Within the diffusive model applicable for Corbino rings, we were able to accurately fit the temperature difference producing the thermopower based on the measured conductance traces and find that it to be consistent with both our experimental temperature estimates and the one derived from independent thermal conductivity data. 
The calculated figure of merit $ZT$ is remarkably high for high magnetic fields indicating that this system is very promising  as a low-temperature cooling device or a heat engine. 

Future work needs to clarify the origin the large voltage signals at the conductance minima, i.e. in the quantized state where both the electric conductance and the thermal conductance values of the Onsager equation vanish. Also different mechanisms might be relevant in these regimes, like temperature driven magnetic flux \cite{dolgo} or temperature dependent contact potentials which cannot equilibrate in the conductance minima \cite{VKcontact}. Another important direction would be the extension of the experiment to lower temperatures. Determining entropy in the fractional quantum Hall regime could answer some urgent questions about the entropy of the suspected non-abelian states.

\section{Acknowledgements}
We thank Klaus von Klitzing for his constant interest and support, Achim G\"uth and Marion Hagel for the wafer lithography, Mirko Lupatini, Luca Alt, and Simon Parolo for help with the experiments. Peter M\"{a}rki provided the amplifiers used in this work while Lars Tiemann contributed the measurement software "Nanomeas" (www.nanomeas.com). We received useful comments on the manuscript from Peter Samuelsson.
We acknowledge support from INTI and CONICET, Argentina.  We are sponsored by PIP-RD 20141216-4905 
of CONICET,  PICT-2017- 2726 and PICT-2018 from Argentina, Swiss National Foundation (Schweizerischer Nationalfonds) NCCR "Quantum
Science and Technology", as well as  the Alexander von Humboldt Foundation, Germany.



\begin{thebibliography}{9}

\bibitem{vK} K. v. Klitzing, G. Dorda, M. Pepper, Phys. Rev. Lett. {\bf 45}, 494, (1980).

\bibitem{fractio} D. C. Tsui, H.L. Stormer, A. C. Gossard, Phys. Rev. Lett. {\bf 48}  1559 (1982).

\bibitem{laughlin} R. B. Laughlin, Phys. Rev. Lett. {\bf 50} 1395 (1983).

\bibitem{jain} J. K. Jain, Adv. Phys. {\bf 41}, 105 (1992).




\bibitem{gerster} T. Gerster, A. M\"uller, L. Freise, D. Reifert, D. Maradan, P. Hinze, T. Weimann, H. Marx, K. Pierz, H. W. Schumacher, F. Hohls, N. Ubbelohde, Metrologia {\bf 56}, 014002  (2018).

\bibitem{topocomp} C. Nayak, S. H. Simon, A. Stern, M. Freedman, S. Das Sarma, Rev. Mod. Phys. {\bf 80}, 1083 (2008).

\bibitem{halpe} K. Yang, B. I. Halperin, Phys. Rev. B {\bf 79}, 115317 (2009).


\bibitem{barlas} Y. Barlas, K. Yang, Phys. Rev B {\bf 85}, 195107 (2012).

\bibitem{Fletcher88} R. Fletcher, M. D'Iorio, W. T. Moore and R. Stoner, J. Phys. C {\bf 21}, 2681 (1988).

\bibitem{Fromhold93} T.M. Fromhold, P.N. Butcher, G. Qin, B.G. Mulimani, J.P. Oxley and B.L. Gallagher, Phys. Rev. B {\bf 48}, 5326 (1993).

\bibitem{therpow8} B. Tieke, U. Zeitler, R. Fletcher, S. A. J. Wiegers, A. K. Geim, J. C. Maan, M. Henini, Phys. Rev. Lett. {\bf 76}, 3630 (1996).

\bibitem{Maximov_2004}S. Maximov, M. Gbordzoe, H. Buhmann, L.W. Molenkamp. D. Reuter, Phys. Rev. B {\bf 70}, 121308 (2004)

\bibitem{dolgo} V. T. Dolgopolov, A. A. Shashkin, N. B. Zhitenev, S. I. Dorozhkin, K. von Klitzing, Phys. Rev. B {\bf 46}, 12560 (1992).

\bibitem{therpow10} H. van Zalinge, R. W. van der Heijden, J. H. Wolter, Phys. Rev. B {\bf 67}, 165311 (2003).

\bibitem{kobaya} S. Kobayakawa, A. Endo, Y. Iye, J. Phys. Soc. Japan {\bf 82}, 053702 (2013).


\bibitem{chick1} W. E. Chickering, J. P. Eisenstein, L. N. Pfeiffer, K. W. West, Phys. Rev. B {\bf 81}, 245319 (2010).

\bibitem{chick2} W. E. Chickering, J. P. Eisenstein, L. N. Pfeiffer, K. W. West, Phys. Rev. B {\bf 87}, 075302 (2013).


\bibitem{Wolfe} J.P. Wolfe, {\it Imaging Phonons: Acoustic Wave Propagation in Solids} (Cambridge: Cambridge University Press, 1998) doi:10.1017/CBO9780511665424

\bibitem{marki2015} P. M\"arki, B. A. Braem, T. Ihn, Rev. Sci. Instrum. {\bf 88}, 085106 (2017).

\bibitem{casati} G. Benenti, G. Casati, K. Saito, R. S. Whitney, Phys. Rep. {\bf 694}, 1 (2017).

\bibitem{jonson} M. Jonson, S. M. Girvin, Phys. Rev. B {\bf 29}, 1939 (1984).









\bibitem{rafa} R. S\'anchez, B. Sothmann, A. N. Jordan, Phys. Rev. Lett. {\bf 114}, 146801 (2015).


\bibitem{janine} S. Kheradsoud, N. Dashti, M. Misiorny, P. P. Potts, J. Splettstoesser, P. Samuelsson, Entropy {\bf 21}, 777 (2019).

\bibitem{daniel} D. Gresta, M. Real, L. Arrachea, Phys. Rev. Lett. {\bf 123}, 186801 (2019).

\bibitem{hof} P. P. Hofer, B. Sothmann, Phys. Rev. B {\bf 91}, 195406 (2015).

\bibitem{peter} P. Samuelsson, S. Kheradsoud, B. Sothmann, Phys. Rev. Lett. {\bf 118}, 256801 (2017).

\bibitem{vanuci} L. Vannucci, F. Ronetti, G. Dolcetto, M. Carrega, M. Sassetti, Phys. Rev. B {\bf 92}, 075446 (2015).

\bibitem{enhan} P. Roura-Bas, L. Arrachea, E. Fradkin, Phys. Rev. B {\bf 97}, 081104(R) (2018).

\bibitem{fabio2} F. Giazotto, F. Taddei, M. Governale, R. Fazio, F. Beltram, New J. Phys. {\bf 9}, 439 (2007).

\bibitem{fu} L. Fu, arXiv:1909.09506.

\bibitem{he}J. He, and T. M. Tritt, Science {\bf 357}, 6358 (2017).

\bibitem{sebas1}A Ozaeta, P Virtanen, FS Bergeret, TT Heikkil\"a,Phys. Rev. Lett. {\bf 112}, 057001 (2014).





\bibitem{VKcontact} K. von Klitzing,  {\it Private Communication} 
 

 

\end{thebibliography}
\end{document}